\documentclass[letter,12pt]{article}
\pdfoutput=1 

\usepackage{jheppub} 

\usepackage[T1]{fontenc} 
\usepackage{amsmath}
\usepackage{amsfonts}
\usepackage{amssymb}
\usepackage{color}

\newcommand{\bea}{\begin{eqnarray}}
\newcommand{\eea}{\end{eqnarray}}

\newcommand{\Sphere}{\mathbb{S}}
\newcommand{\C}{\mathbb{C}}

\newcommand{\D}{\mathcal{D}}

\title{KdV-charged black holes}

\author[a,b]{Anatoly Dymarsky,} 
\author[a]{and Sotaro Sugishita}

\affiliation[a]{University of Kentucky,\\Lexington, KY, USA 40506\\}
\affiliation[b]{Skolkovo Institute of Science and Technology,\\Skolkovo Innovation Center, Moscow, Russia\\}

\emailAdd{a.dymarsky@uky.edu}
\emailAdd{sotaro.s@uky.edu}

\abstract{
We construct black hole geometries in AdS$_3$  with non-trivial values of  KdV charges. The black holes  are holographically  dual to quantum KdV Generalized Gibbs Ensemble in  2d CFT. 
They satisfy thermodynamic identity and thus are saddle point configurations of the Euclidean gravity path integral. We discuss holographic calculation of the KdV generalized partition function 
and show that for a certain value of chemical potentials new geometries, not the conventional BTZ ones, are the leading saddles. 
}

\begin{document} 
\maketitle
\flushbottom

\section{Introduction}
\label{sec:intro}
Thermalization and non-equilibrium dynamics in two-dimensional conformal field theory is a rich subject touching on many topics  of contemporary interest, from cold atom experiments to chaos in quantum gravity  \cite{calabrese2006time,calabrese2007quantum,Roberts:2014ifa}.
Dynamics of 2d CFTs  is constrained by presence of an infinite tower of local conserved quantum KdV charges $\hat{Q}_{2k+1}$, which commute with each other and the CFT Hamiltonian  $\hat{Q}_1=L_0-c/24$. The qKdV integrable structure of 2d CFTs has been actively studied in the past \cite{bazhanov1996integrable,bazhanov1997integrable,bazhanov1999integrable,bazhanov97zero}, as well as more recently \cite{maloney2018thermal,Kotousov:2019nvt,LeFloch:2019wlf}. 

Interest in quantum KdV charges was rekindled recently in the context of generalized thermalization of quantum integrable systems and  Generalized Gibbs Ensemble, which is expected to  emerge as a result of thermalization dynamics \cite{vidmar2016generalized}. In particular, if the initial 2d CFT state carries non-trivial qKdV charges, local physics at late times  was argued to be given by the KdV GGE state \cite{cardy2016quantum},
\bea
\label{GGEi}
\rho_{\rm GGE}=e^{-\sum_k \tilde{\mu}_k \hat{Q}_{2k+1}}/{\mathcal Z},
\eea
where 
\bea
{\mathcal Z}={\rm Tr}\,e^{-\sum_k \tilde{\mu}_k \hat{Q}_{2k+1}} \label{Z}
\eea
is the KdV generalized partition function. Thermodynamics and other basic properties of the KdV GGE are not well understood. In fact $\mathcal Z$
is not known explicitly  even for simplest theories. This paper is paving the way for calculation of ${\mathcal Z}$ in the large $c$ limit. In the previous works on the subject, both on holographic and the CFT sides \cite{de2016remarks,maloney2018generalized,GGE,GGE2}, it was implicitly assumed that the BTZ black holes, i.e.~eigenstates of $\hat{Q}_1=L_0-c/24$ on the CFT side, are the  leading saddle point configurations contributing to the KdV generalized partition function \eqref{Z}. Provided this is the case, $1/c$ corrections can be calculated by quantizing small  fluctuations near the BTZ saddle, as was done on the CFT side in  \cite{GGE2}. We show this is not always the case, namely there are novel black hole configurations, which correspond to complicated CFT states, not $\hat{Q}_1$ eigenstates, which are leading contributions to ${\mathcal Z}$, at least for the particular values of chemical potentials $\tilde{\mu}_{2k+1}$.   

These KdV-charged black holes, which we construct explicitly, are gravity dual to the KdV GGE state \eqref{GGEi}. Conventional BTZ geometries emerge as  a particular case, which is  dual to the conventional Gibbs Ensemble, i.e.~when all $\tilde{\mu}_{2k+1}=0$ except for $\tilde{\mu}_1=\beta$.

The mapping between a particular KdV-charged black hole and \eqref{GGEi} is non-trivial. Namely, the averaged values of $\hat{Q}_{2k+1}$ in $\rho_{\rm GGE}$ should match those in the holographic configuration. Thus, at infinite $c$, corresponding classical black hole geometry analytically continued to  Euclidean signature should be a leading saddle point configuration of the corresponding gravity path integral evaluating $\mathcal Z$. To support the validity of this prescription,  we explicitly show for a particular simple KdV-charged configuration that for the certain values of $\tilde{\mu}_{2k+1}$ its contribution exceeds those of the BTZ black holes. 

We provide a general proof that the KdV-charged black holes satisfy the first law of thermodynamics. This and other properties of the geometries follow from the integrable structure of the KdV equation and its relation to the co-adjoint orbit of Virasoro algebra. To make the presentation self-contained, we start with a succinct  introduction of mathematical preliminaries in the next section. After that, in section \ref{sec:qKdV}, we explain how classical integrability of KdV equation gives rise to quantum KdV charges in 2d CFTs. The new geometrical solutions are constructed and analyzed in section \ref{sec:gravity}, where we also prove the thermodynamic identity. Section \ref{sec:CFT} discusses dual field theory interpretation of the new geometries. In section \ref{sec:Z} we calculate KdV generalized partition function $\mathcal Z$ on the gravity side in the case when only $\tilde{\mu}_1$, and $\tilde{\mu}_3$ are non-zero, and see that only thermal AdS and BTZ configurations contribute. Then in section \ref{sec:5} we turn on $\tilde{\mu}_5$ and see that new KdV-charged black hole configurations appear and may become leading for a particular range of parameters. 
We conclude with a discussion in section \ref{sec:Discussion}.

\section{Mathematical preliminaries}
In this section we provide mathematical preliminaries necessary for the general discussion of the consecutive sections. We aimed at a self-contained but concise presentation and many details and proofs were omitted. The reader is advised to consult the original papers by Witten, Novikov, and others \cite{Witten:1987ty,novikov1974periodic,lazutkin1975normal,dubrovin1974periodicity,novikov1984theory}
for a systematic presentation of the geometry of the co-adjoint orbits of  Virasoro algebra, finite-zone solutions of the generalized KdV equations, and other related questions. 

\subsection{Co-adjoint orbit of Virasoro algebra}
\label{sec:orbit}
We start by introducing the group ${\rm diff}\,\Sphere^1$ of diffeomorphisms of a circle. Elements of  ${\rm diff}\,\Sphere^1$ are  monotonically increasing functions $\tilde{\varphi}=g(\varphi)$,
\bea
g(2\pi)=g(0)+2\pi,
\eea
such that $g$ is an invertible map of a circle into itself, $g(\varphi)=g(\varphi') \Rightarrow \varphi=\varphi'$.
Corresponding Lie algebra is the Witt algebra of vector fields on a circle $f(\varphi)\partial_\varphi$. 

Next we consider a periodic ``potential''  $u(\varphi)$, $u(\varphi+2\pi)=u(\varphi)$ and a ``wave-function''
$\psi(\varphi)$ satisfying ``Schr$\ddot{\rm o}$dinger'' equation (properly called Hill's equation),
\bea
\label{Hill}
-\psi''+{u\over 4}\, \psi=0.
\eea
Diffeomorphisms $g \in {\rm diff}\,\Sphere^1$ naturally act on $u$ and $\psi$,
\bea
g:&& \psi(\varphi)\rightarrow \tilde{\psi}(\tilde{\varphi}),\\ 
g:&& u(\varphi)\rightarrow \tilde{u}(\tilde{\varphi}), 
\eea
such that the Hill's equation continue being satisfied (the derivative is with respect to $\tilde{\varphi}$),
\bea
-\tilde{\psi}''+{\tilde{u}\over 4}\, \tilde{\psi}=0.
\eea
The new potential and the new wave-function are defined via
\bea
\label{tildepsi}
\tilde{\psi}(\tilde{\varphi}(\varphi))&=&\psi(\varphi) \left({d\tilde{\varphi}\over d\varphi}\right)^{1/2},\\
\label{tildeu}
\tilde{u}(\tilde{\varphi}(\varphi))&=& \left({d\tilde{\varphi}\over d\varphi}\right)^{-2}\left[u(\varphi)+2\{\tilde{\varphi},\varphi\}\right].
\eea
Here for any $\theta(\varphi)$
\bea
\{\theta,\varphi\}\equiv  {\theta'''\over \theta'}-{3\over 2}\left({\theta''\over \theta'}\right)^2,
\eea
is the Schwarzian derivative. 

An infinitesimal transformation 
\bea
g(\varphi)=\varphi-\epsilon f(\varphi)
\eea
acts on the potential as follows, 
\bea
\tilde{u}(\varphi)=u(\varphi)+ \epsilon\, \D f,\qquad \D\equiv (\partial u)+ 2  u\partial -2\,\partial^3.
\eea
As we will now see, this is the action of Virasoro algebra, central extension of Witt algebra, on its co-adjoint orbit. 

Elements of Virasoro algebra are the pairs $(f,a)$ where $f$ is a vector field and $a$ is a $\C$-number with the following commutation relation 
\bea
\label{Lie}
[(f_1,a_1),(f_2,a_2)]=(f_1 f_2'-f_1' f_2,a),\qquad a=\int_0^{2\pi} d\varphi (f_1''' f_2-f_1 f_2''').
\eea
Co-adjoint space is the linear space dual to the algebra. Its elements are the pairs $[u,\hat{t}]$ where $u(\varphi)d\varphi^2$ is a ``two-differential'' and $[0,\hat{t}]$ is an element formally dual to $(0,1)$. We want  $\hat{t}$ to be common for all elements, and therefore we can reduce the notations from $[u,\hat{t}]$ to simply $u$, such that the scalar product is 
\bea
\langle (f,a),u\rangle=a+\int_0^{2\pi}d\varphi\, u f.
\eea 
It is easy to see that $\langle (f,a),u\rangle$ is invariant under the action of a Virasoro algebra element $(v,b)$ provided,
\bea
\label{Va}
\delta f=v f'-v' f,\quad \delta a=\int_0^{2\pi} d\varphi (v''' f-v f'''),\quad 
\delta u=\D v. 
\eea

Action of ${{\rm diff}\,\Sphere^1}$  \eqref{Va} foliates the space of all $u(\varphi)$ into orbits -- the co-adjoint orbits of Virasoro algebra. Starting with some potential $u$ one defines a sub-algebra of stabilizers $f$ of $u$ such that 
\bea
\delta u=\D f=0.
\eea
In full generality there could be either one or three linearly independent stabilizers \cite{Witten:1987ty}, which must be closed in the Lie algebra sense. Then the orbit is defined by the action of all possible diffeomorphisms $g(\varphi)$ on the given $u$, modulo the stabilizer subgroup. The simplest orbit is obtained 
starting from a constant $u(\varphi)=u_0$. In this case the stabilizer is unique, $f=1$, up to an overall rescaling, with an exception of the case when $u_0\neq -n^2$ for some integer $n$. These are the orbits ${\rm diff}\,\Sphere^1/\Sphere^1$ in the notations of Witten \cite{Witten:1987ty}, or stable orbits in the notations of Lazutkin and Pankratova \cite{lazutkin1975normal}.
Quantization of such an orbit gives rise to Verma module. 

When  $u(\varphi)$ belongs to an orbit ${\rm diff}\, \Sphere^1/\Sphere^1$  the stabilizer vector field $f$ for each $u$ is unique and sign-definite. The converse is also correct and easy to see. Let us consider $f$, such that it is sign-definite and $\D f=0$. We first notice that 
\bea
2\pi f_0^{-1}=\int_0^{2\pi} {d\varphi\over f},
\eea
is invariant under the diffeomorphism ${\rm diff}\, \Sphere^1$, as follows straightforwardly from \eqref{Lie}. Next, one  can define the diffeomorphism $\varphi\rightarrow \tilde{\varphi}=g(\varphi)$,
\bea
\label{fdiff}
d\tilde{\varphi}=f_0{d\varphi\over f},
\eea
which brings $f$ to a constant form $f_0$. This is the diffeomorphism which  brings $u(\varphi)$ to a constant, as follows from applying \eqref{tildeu},
\bea
\label{u0}
\tilde{u}(\tilde{\varphi})=u_0={u f^2+f'^2-2f f''\over f_0^2}.
\eea
That $u_0$ is a constant can be verified by differentiating it,  $\tilde{u}'=(f/f_0) \partial_\varphi u_0=(f^2/f_0^3) \D f=0$. An alternative way to obtain the same expression is to start with $\D f=0$ and solve it as an equation for $u$, 
\bea
\label{sol}
u(\varphi)={u_0 f_0^2-f'^2+2f f''\over f^2}.
\eea
Here $u_0 f_0^2$ appears as an integration constant. It is straightforward to see that 
 \eqref{sol}  is compatible with \eqref{Va} only if $u_0 f_0^2$ is invariant under the diffeomorphisms. Hence $u_0 f_0^2$  is equal to $u f^2$ when $u(\varphi)$ is $\varphi$-independent and hence so is $f$. 
Finally, we note that $u_0$ is the only invariant characterizing the orbit, and its invariance under the diffeomorphisms follows straightforwardly from \eqref{u0} and \eqref{Lie}.

The space of all potentials is a Poisson manifold with the Poisson bracket \cite{magri1978simple},
\bea
\label{P}
{c\over 24}\{u(\varphi_1),u(\varphi_2)\}=-2\pi \D\delta(\varphi_1-\varphi_2),
\eea
where $c$ is some numerical parameter. 
Written  in terms of the Fourier series 
\bea
\label{Fourier}
{c\over 24}(u(\varphi)+1)=\sum_k \ell_k e^{ik\varphi},
\eea the Poisson brackets \eqref{P} reduce to Virasoro algebra
\bea
\label{Vir}
i\{\ell_n,\ell_m\}=(n-m)\ell_{n+m}+{c(n^3-n)\over 12}\delta_{n+m}.
\eea
In particular for any functional ${\mathcal H}[u(\varphi)]$,
\bea
\label{Hflow}
{c\over 24}\{{\mathcal H},u(x)\}=\D f,\qquad f=2\pi {\delta {\mathcal H}\over \delta u(x)}.
\eea
Here $\D f$ is  as the Hamiltonian vector field associated with $\mathcal H$ in the space of potentials $u(\varphi)$.

Since the Hamiltonian vector field \eqref{Hflow} has the form of \eqref{Va} with some appropriate $v=f$, Hamiltonian flow does not move $u(x)$ away from the orbit, hence on the space of all potentials the Poisson bracket is degenerate. Restricting it to a particular orbit removes this degeneracy, and \eqref{P} defines a symplectic form, 
such that each orbit is a symplectic manifold. This symplectic form is the Kirillov-Kostant form on the co-adjoint orbit of Virasoro algebra \cite{gervais1982dual}, as is also evident from \eqref{Vir}.

\subsection{KdV hierarchy}
We now go back to Hill's equation \eqref{Hill} and  extend it to a full Schr$\ddot{\rm o}$dinger eigenvalue problem (Sturm-Liouville equation),
\bea
\label{Sch}
-\psi''+{u\over 4}\psi=\lambda\, \psi.
\eea
The (non-degenerate) eigenvalues of periodic $\psi(2\pi)=\psi(0)$ and anti-periodic $\psi(2\pi)=-\psi(0)$ problems constitute the so-called spectral data of $u(\varphi)$. Different potentials may share the same spectral data. In fact there is an infinite family of infinitesimal deformations which are isospectral, i.e.~preserve the spectral data.  The isospectral deformations are generated by the Hamiltonian flow  associated with the Poisson bracket \eqref{P},
\bea
\delta u={c\over 24}\{Q_{2k-1},u\},
\eea
where $Q_{2k-1}$ are the so-called KdV generators, which can be defined iteratively,
\bea
\label{Qkdv}
Q_{2k-1}={1\over 2\pi}\int_0^{2\pi}d\varphi\, R_k,\qquad 
\partial R_{k+1}={k+1\over 2k+1}\D R_k,\qquad R_0=1.
\eea
The Gelfand-Dikii polynomials $R_k$ \cite{gel1975asymptotic} satisfy various relations, in particular 
\bea
{c\over 24}\{Q_{2k-1},u\}=(2k-1)\partial R_k.
\eea
First few  $R_k$ and $Q_{2k-1}$ are given by 
\bea
R_0&=&1,\quad R_1=u,\quad R_2=u^2-{4\over 3}\partial^2u,\quad R_3=u^3-4 u\partial^2 u-2(\partial u)^2+{8\over 5}\partial^4 u,\qquad \\
I_0&=&u,\quad I_1={u^2},\quad I_2={u^3+2(\partial u)^2},
\quad I_3={u^4}+8u(\partial u)^2+{16(\partial^2 u)^2\over 5},
\eea
where $2\pi\, Q_{2k+1}=\int_0^{2\pi}d\varphi\, I_k$. Of course $I_k$ and $R_{k+1}$  differ only by a full derivative.

The name KdV comes from the form of the flow generated by $Q_3$. Assuming it defines a $t$-dependent function $u(t,x)$ via
\bea
\dot{u}={c\over 24}\{Q_{3},u\}=6 u u'-{4} u''',
\eea
we immediately recognize the original KdV equation (perhaps up to a notational difference). 

The KdV charges are in involution, $\{Q_{2k-1},Q_{2l-1}\}=0$, yet the action of $Q_{2k-1}$ on a given $u(x)$ is usually non-trivial. It is known that the corresponding Hamiltonian flows exhaust all possible  isospectral deformations. 

\subsection{Finite-zone ``Novikov'' solutions}
\label{sec:Novikov}
For an arbitrary complex $\lambda$ equation \eqref{Sch} has two solutions, which can be combined into a complex-valued quasi-periodic wave-function
\bea
\psi(\varphi+2\pi)=\psi(\varphi)e^{2\pi i p(\lambda)}.
\eea
For real $\lambda$ the quasi-momentum $p(\lambda)$ is either real or pure imaginary. In the latter case $\lambda$ belongs to the so-called forbidden zone. Forbidden zones stretch between two consecutive eigenvalues of periodic or anti-periodic problem. The zone disappears if the periodic or anti-periodic problem is double degenerate. The quasi-momentum is a complex function with the branch-cuts along the forbidden zones and $\lambda\leq \lambda_0$, where $\lambda_0$ is the energy of the ground state.  For example all eigenvalue of the periodic and antiperiodic problems for the constant potential $u=u_0$ are double degenerate (except for the ground state), 
\bea
\label{const}
\lambda^{(n)}={n^2+u_0\over 4},\qquad n\geq 0.
\eea
Therefore there are no forbidden zones and  $p(\lambda)=\sqrt{\lambda-u_0/4}$.

A special class of potentials with only a finite number of degeneracies lifted, and hence only a finite number of forbidden zones, are called finite-zone potentials. For example a one-zone potential will have a double-degenerate eigenvalue $\lambda^{(k)}$ for some $k\geq 1$ split into two, $\lambda^{(k)}_-=\lambda_1$ and $\lambda^{(k)}_+=\lambda_2$, while all other eigenvalues of periodic and antiperiodic problems remain double-degenerate (although their values are no longer  given by \eqref{const}). It turns out that the values of all double-degenerate eigenvalues are uniquely fixed by the vacuum energy $\lambda_0$ and the ends of the zones, which in our case are  $\lambda_1,\lambda_2$. Corresponding $p(\lambda)$ has two branch-cuts from $-\infty$ to $\lambda_0$ and from $\lambda_1$ to $\lambda_2$ and is given by an Elliptic integral discussed below. It is naturally defined on a torus, a Riemann curve of genus $1$.
More generally a finite zone potential is specified by $\lambda_i$, $0\leq i\leq 2n$ and is defined on a hyperelliptic curve of genus $n$. 

Finite-zone potentials emerge as solutions of the static generalized KdV equation  \cite{novikov1974periodic},
\bea
\label{N}
\{ {\mathcal H},u\}=0,\qquad {\mathcal H}=\sum_{i=0}^{n} \mu_{2i+1}Q_{2i+1},\qquad \mu_{2n+1}\neq 0.
\eea
In fact the following is true. Any solution of \eqref{N} is a $m\leq n$-zone potential, and all $n$-zone potentials can be obtained from \eqref{N} with the appropriate $\mu_{2i+1}$. 

For the given spectral data  specified by the ends of the zones $\lambda_i$, $0\leq i\leq 2n$, quasi-momentum is 
specified indirectly by its differential 
\bea
\label{dp}
dp={\lambda^n+r_{n-1}\lambda^{n-1}+\dots r_0\over 2 y}d\lambda,
\eea
which is defined on the Riemann curve 
\bea
\label{curve}
y^2=\prod_{i=0}^{2n}(\lambda-\lambda_i).
\eea
Coefficients $r_i$  are fixed by the condition that 
\bea
\label{zero}
\oint\limits_{a_i} dp=2\int_{\lambda_{2i-1}}^{\lambda_{2i}} dp =0,
\eea
vanishes for any a-cycle, defined as the brunch-cuts of $y$ from $\lambda_{2i-1}$ to $\lambda_{2i}$. Because of \eqref{zero}, function $p(\lambda)$ defined such that $dp=({\partial p}/\partial \lambda)d\lambda$ is a well defined function on the Riemann curve \eqref{curve}. 

As a complex function $p(\lambda)$ has branch-cuts along the forbidden zones, and therefore finite zone solutions are also called finite- or multi-cut solutions, the language we occasionally use in this paper.

Each Hamiltonian flow generated by $Q_{2k+1}$ is isospectral, hence it deforms a finite-zone solution  into another, such that $\{{\mathcal H},u\}=0$ continue being  satisfied. For any fixed $Q_{2k-1}$, $1\leq  k\leq n+1$, values of all higher charges $Q_{2k-1}$, $k>n+1$ are fixed and the space of solutions is an $n$-dimensional torus (Jacobian of the hyperelliptic curve \eqref{curve}). All charges $Q_{2k+1}$ generate a flow on the Jacobian, which is ergodic in a general case. The exception being the flow generated by $Q_{1}$ which is equivalent to the shift $\varphi\rightarrow \varphi +{\rm const}$ and therefore $2\pi$-periodic. 

Spectral data is not invariant under ${\rm diff}\, \Sphere^1$ transformations \eqref{tildeu}, but $\lambda=0$ quasi-periodic eigenfunction $\psi$ transforms into $\tilde{\psi}$ according to \eqref{tildepsi}, which is different only by an overall $2\pi$-periodic factor. Hence $e^{2\pi i p(0)}$ is an invariant on the whole co-adjoint orbit. Considering a constant representative  $u(\varphi)=u_0$ yields 
\bea
u_0=-4 p^2(0). \label{h}
\eea

\subsection{Example: one-cut solutions}
\label{sec:1zone}
In this section we solve the generalized KdV equation \eqref{N} for $n=1$,
\bea
\frac{c}{24}\{Q_3+\alpha\, Q_1,u\}=6u\, u'- 4 u'''+\alpha\, u'=0.
\eea
By integrating this equation twice we obtain an effective problem for a particle moving in a cubic potential 
\bea
\label{potential}
{u'^2\over 2} +V(u)=E, \quad 2(E-V)={1\over 2}(u-u_1)(u-u_2)(u-u_3),\\
 s1:=u_1+u_2+u_3=-\alpha/2. \label{constraint}
\eea
Two out  of three  parameters  $u_i$ are free. They specify the values of $Q_1,Q_3$ evaluated on the solution. From \eqref{potential}
$u(\varphi)$ can be obtained easily in terms of Weierstrass's elliptic function $\wp$  by specifying the initial  condition $u(\varphi_0)$. We require the solution to be $2\pi$-periodic, which imposes a condition on ``energy'' $E$, leaving only one free parameter, besides $\varphi_0$. There is in fact an infinite tower of solutions with the period $2\pi/k$ for positive integer $k$, each being parametrized by one continuous parameter, in addition to $\varphi_0$. Weierstrass's function is associated with a torus, and we choose $s_1$ and torus  modular parameter $q=e^{i\pi \tau}$ as the independent parameter of one-cute solution,
\bea
\label{def:u1}
u_1={s_1\over 3} -{2k^2\over 3}\left(\theta_2(0;q)^4+\theta_3(0;q)^4\right),\\
u_2={s_1\over 3} +{2k^2\over 3}\left(\theta_2(0;q)^4-\theta_4(0;q)^4\right),\\
u_3={s_1\over 3} +{2k^2\over 3}\left(\theta_3(0;q)^4+\theta_4(0;q)^4\right).
\eea 
We order $u_1\leq u_2\leq u_3$, such that the periodic solution describes the oscillations of a ``particle'' between $u_1$ and $u_2$. Sometimes instead of  $u_i$ it is convenient to use $s_1$ and 
\bea
\label{def:s2}
s_2&:=&u_1 u_2+u_2 u_3+u_1 u_3=-{k^4\over \pi^4}g_2(\tau)+{s_1^2\over 3},\\
s_3&:=&u_1 u_2 u_3= -{2k^6\over \pi^6}g_3(\tau)-{k^4\over 3\pi^4}s_1 g_2(\tau)+{s_1^3\over 27}.
\label{def:s3}
\eea
Here $g_2$ and $g_3$ are modular forms. 
The value of $Q_1$ can be written in terms of $u_i$ as
\begin{align}
\label{Q1_onecut}
Q_1=u_3+(u_2-u_3)\frac{{}_2F^{}_1\left(\frac{3}{2},\frac{1}{2},1;\frac{u_2-u_1}{u_3-u_1}\right)}{{}_2F^{}_1\left(\frac{1}{2},\frac{1}{2},1;\frac{u_2-u_1}{u_3-u_1}\right)}.
\end{align}
Higher KdV charges can be expressed through $Q_1$ and $s_i$,
\begin{align}
\label{Q3Q5_onecut}
Q_3=\frac{1}{3}(2s_1 Q_1 -s_2), 
\quad
Q_5=\frac{2s_1^2-s_2}{5}Q_1-\frac{s_1 s_2 +s_3}{5}.
\end{align} 

In terms of the spectral data, one-zone potential is characterized by three eigenvalues of the Schr$\ddot{\rm o}$dinger equation \eqref{Sch}, the ground state  $\lambda_0$, and the ends of the forbidden zone, $\lambda_1,\lambda_2$,
\bea
\label{lambdau}
\lambda_0={u_1+u_2\over 8},\quad \lambda_1={u_1+u_3\over 8},\quad \lambda_2={u_2+u_3\over 8}.
\eea
When the ``energy'' $E$ is small,  meaning $E-V$ approaches zero, the ``particle'' oscillates near the local minimum of the potential with the period $2\pi/k$, and the values of $\lambda_1$  and $\lambda_2$ approach $\lambda_0+{k^2/4}$ from both sides. This corresponds to a small perturbation of the constant potential which removes degeneracy of just one eigenvalue in  \eqref{const}.

Besides one-cut solutions with non-constant $u(\varphi)$ there are also two $\varphi$-independent solutions of \eqref{potential} corresponding to a ``particle'' sitting at the top or bottom of the potential.

\section{qKdV symmetry in CFT$_2$}
\label{sec:qKdV}
The co-adjoint orbit of Virasoro algebra ${\rm diff}\, \Sphere^1/\Sphere^1$ is a symplectic manifold with the non-degenerate Poisson bracket \eqref{P}. Upon quantization, it gives rise to Verma module with the primary (highest weight) state $|\Delta\rangle$ of dimension 
\bea
\label{spectrum}
\Delta={c\over 24}(u_0+1).
\eea
In the classical case $u(\varphi)$, or equivalently its Fourier modes $\ell_n$ \eqref{Fourier}, subject to a constraint which ensures $u(\varphi)$ belongs to the orbit, are the coordinates on the orbit. Upon quantization they become Virasoro algebra generators $L_n$, while $u$ becomes  stress-energy tensor in a CFT$_2$  on a cylinder with the central charge $c$,
\bea
T={c\over 24}u.
\eea
It is then easy to recognize  \eqref{tildeu} as the standard expression for the change of stress-energy tensor upon a coordinate transformation. 

The Poisson brackets \eqref{P} were originally introduced  in the context of higher KdV equations  \cite{magri1978simple}, and soon the connection with the Virasoro algebra was noticed by Gervais and Neveu \cite{gervais1982dual}. Later Gervais suggested that classical KdV charges \eqref{Qkdv}, upon quantization, should give rise to mutually-commuting quantum operators \cite{gervais1985infinite,gervais1985transport}. While being very intuitive, this proposal is not trivial. Since the higher generators are non-linear in $u$, their quantum counterparts will depend on the normal ordering and may no longer commute as a result. This question was fully resolved only in \cite{bazhanov1996integrable,bazhanov1997integrable,bazhanov1999integrable} where existence of an infinite tower of local commuting qKdV charges $\hat{Q}_{2k+1}$ was established. Their definition, besides normal ordering, is also explicitly $c$-dependent. Thus, for example, first few charges  in terms of the stress-tensor are 
\bea
\nonumber
\hat{Q}_{1}={1\over 2\pi}\int_0^{2\pi} d\varphi\, T,\quad \hat{Q}_{3}={1\over 2\pi}\int_0^{2\pi} d\varphi (TT),\quad \hat{Q}_{5}={1\over 2\pi}\int_0^{2\pi} d\varphi\, (T(TT))+{c+2\over 12}(\partial T)^2.
\eea
In our notations classical charges $Q_{2k+1}$ give rise to $\left({c\over 24}\right)^{-k-1}\hat{Q}_{2k+1}$.
Notice however that $\left({c\over 24}\right)^3 Q_5$ is {\it not} equal to  $\hat{Q}_5$  upon substitution $u\rightarrow {24\over c}T$ and normal ordering. An extra term $(\partial T)^2/6$ is necessary to assure commutativity. 
 
In terms of Virasoro algebra generators $\hat{Q}_1$ is simply the CFT Hamiltonian $L_0-{c\over 24}$. Expressions for $\hat{Q}_3,\hat{Q}_5$ in terms of $L_n$ are also known \cite{bazhanov1996integrable}, as well as for $\hat{Q}_7$ \cite{dymarsky2019zero}, but quickly become prohibitively complicated.

The gravity configurations discussed below are dual to the GGE state 
\bea
\label{GGE}
\rho \propto e^{-\sum_k \tilde{\mu}_{2k+1}\hat{Q}_{2k+1}},
\eea
which can be understood as a state in the original CFT with the Hamiltonian $H=\hat{Q}_1=L_0-c/24$, as well as a state in a theory with the KdV-deformed Hamiltonian 
\bea
\label{Hdef}
H=\sum_k \tilde{\mu}_{2k+1}\hat{Q}_{2k+1}.
\eea
Since all $\hat{Q}_{2k+1}$ commute, vacuum of the original theory $|0\rangle$ is an eigenstate of  $H$, but may not be the ground state for some particular choice of 
$\tilde{\mu}_k$.

\section{New black hole geometries}
\label{sec:gravity}
In this section we construct the black hole geometries in pure gravity in AdS$_3$ with the KdV-deformed  boundary conditions \cite{perez2016boundary,Fuentealba:2017omf,Ojeda:2019xih}, which is a gravity dual theory for the 2d CFT with the deformed Hamiltonian \eqref{Hdef}.  

\subsection{Gravity in AdS$_3$ with the deformed boundary conditions}
Pure gravity in AdS$_3$ has no local degrees of freedom and the geometry is fixed by the behavior at the boundary.  
Provided the  boundary is parametrized by an angular variable $\varphi$ and time $t$, metric is fixed in terms of two pairs of functions $u(t,\varphi),f(t,\varphi)$ and  $\bar{u}(t,\varphi),\bar{f}(t,\varphi)$, 
\begin{eqnarray}
g_{tt}&=& - f\bar{f} r^2+ \frac{\ell^2}{4}[(f' - \bar{f}')^2+f(f u -2f'')+\bar{f}(\bar{f} \bar{u} -2\bar{f}'')]
\\ \nonumber
&& \quad-\frac{\ell^4}{16r^2}(f u -2f'')(\bar{f} \bar{u} -2\bar{f}''),
\\
g_{tr}&=&
-\frac{\ell^2}{2 r}(f'-\bar{f}'),
\\
g_{t\varphi}&=& \frac{r^2}{2}(f-\bar{f})+\frac{\ell^2}{4}(f u -\bar{f} \bar{u} -f''+\bar{f}'')
\\ \nonumber
&&\quad+\frac{\ell^4}{32 r^2}\left[\bar{u}(f u -2f'')-u(\bar{f}\bar{u} -2\bar{f}'')\right],\\
g_{rr}&=& \frac{\ell^2}{r^2}, \\
g_{r\varphi}&=&0,\\
g_{\varphi\varphi}&=&\left(r+\frac{\ell^2}{4r}u\right)\left(r+\frac{\ell^2}{4r}\bar{u}\right),
\end{eqnarray}
subject to a constraint 
\bea
\label{uf}
\dot{u}=\D f,\qquad \dot{\bar{u}}=-\bar{\D} \bar{f}.
\eea
Here $\ell$ is the radius of AdS$_3$, $\dot{}\, $ stands for $t$-derivative and $'$ for $\varphi$-derivative. 
There is freedom in choosing boundary conditions connecting $f$ and $u$ (and similarly for $\bar{u},\bar{f}$), which corresponds to choosing different Hamiltonians in dual CFT. The choice $f=1$ is conventional and corresponds to the conventional CFT Hamiltonian $H=L_0-{c/24}$ \cite{Brown:1986nw,Bunster:2014mua}.  Following \cite{perez2016boundary} we consider the KdV boundary conditions 
\bea
\label{fu}
f=2\pi {\delta {\mathcal H}\over \delta u},
\eea
where ${\mathcal H}$ is some linear combination of  KdV charges \eqref{Qkdv},
\bea
\label{AdSH}
{\mathcal H}=\sum_{i=0} \mu_{2i+1}Q_{2i+1}.
\eea
Combining \eqref{uf} and \eqref{fu} we can rewrite boundary equations of motion as follows 
\bea
\label{EOM}
\dot{u}={c\over 24}\{{\mathcal H},u\}.
\eea

At this point the connection with the dual CFT becomes apparent:  $u$ is the holographic dual of the stress tensor,
\bea
u={24\over c}T,\qquad \bar{u}={24\over c}\bar{T},
\eea
and the CFT Hamiltonian is a linear combination of quantum KdV charges 
\bea
\label{H}
H=\sum\limits_{i=0} \tilde{\mu}_{2i+1}\, \hat{Q}_{2i+1},\quad \tilde{\mu}_{2k+1}=\left({24\over c}\right)^{k+1}\mu_{2k+1}, 
\eea and similarly for $\bar{H}$. We do not assume here that the considered theory of gravity is pure (quantum) gravity in AdS$_3$ and dual CFT is a particular (hypothetical) dual theory. Rather we work in the large $c$ limit when matter fields in the bulk may be present but do not back-react at the leading order.

\subsection{KdV-charged black holes}
\label{sec:BH}
In what follows we assume that theory and the geometry (dual state) are symmetric under the exchange of left and right sectors,  $H=\bar{H}$, $u=\bar{u}$, 
which automatically implies that the geometry is static $\dot{u}=\D f=0$, $f=2\pi {\delta {\mathcal H}\over du}$. In terms of the section \ref{sec:orbit} this means $f$ is the stabilizer of $u$, hence in most cases it can be reconstructed from $u$ uniquely up to an overall multiplication.
From now on we can consider $u$ from the left sector only. The metric reduces to 
\begin{eqnarray}
\label{met1}
    g_{tt}&=&-\left(f r-{\ell^2\over 4r}(u f-2f'')\right)^2,\\
    g_{\varphi\varphi}&=&\left(r+{\ell^2\over 4r}u\right)^2
    \label{met2},\qquad 
    g_{rr}={\ell^2\over r^2},
\end{eqnarray}
while all other components vanish. There are two different cases to consider. If the spatial circle parametrized by  $\varphi$  is shrinkable, that fixes $u=-1$ to avoid conical singularity. This is the pure AdS$_3$ geometry, which is dual in the holographic sense to CFT with the Hamiltonian \eqref{H} in the vacuum state $|0\rangle$, which is the ground state of the original CFT with the undeformed Hamiltonian $H=\hat{Q}_1=L_0-c/24$.

In a general case $\varphi$-circle is not shrinkable which indicates the geometry has a horizon located at $r^2=r_h^2(\varphi)\equiv \ell^2(u-2f''/f)/4$, where $g_{tt}$ vanishes. This is a black hole solution which carries non-trivial  KdV-charges, the ``KdV-charged'' black hole. If $u$ is a constant, the geometry reduces to the conventional BTZ black hole geometry \cite{banados1992black}, albeit in a theory of gravity with different boundary conditions. If the sum in \eqref{AdSH}  is finite, corresponding $u(\varphi)$ is a finite-zone Novikov solution described in the section \ref{sec:Novikov}. 

 It is well known that any pure gravity solution in AdS$_3$ with a non-shrinkable spatial circle must be diffeomorphic to  BTZ geometry. Therefore the geometry \eqref{met1} must possess two Killing vectors which we will readily identify. Since the solution is static, one Killing vector is simply $\partial_t$.  Another one can be found by solving the Lie derivative equation $\nabla_\mu \xi_\nu+\nabla_\nu \xi_\mu=0$, yielding
\bea
\label{Kxi}
 \xi^t=0,\quad \xi^r=-f'(\varphi)r,\quad \xi^\varphi=f-{2\ell^2 f''\over {\ell^2 u+4r^2}}.
\eea
We can parametrize the trajectory along the Killing vector with a parameter $\tilde{\varphi}$, 
\bea
\label{phiphi}
d\varphi=a \left(f-{2\ell^2 f''\over {\ell^2 u+4r^2}}\right)d\tilde{\varphi}.
\eea
The constant $a$ here is introduced to ensure that $\tilde{\varphi}$ is $2\pi$-periodic. At this point we assume that $f(\varphi)$ is sign-definite. By taking $r$ to infinity we readily find 
\bea
2\pi a=\int_0^{2\pi} {d\varphi \over f}={2\pi \over f_0}.
\eea
By comparing this with the general case discussed in the section \ref{sec:orbit} we readily find that $f$ is the vector field which is a stabilizer of $u$, $f_0$ is invariant under coordinate transformations of the circle, and the Killing vector \eqref{Kxi} maps the original geometry to the BTZ geometry with a constant $u=u_0$ given by \eqref{u0}. The transformation of the boundary variable $\varphi\rightarrow \tilde{\varphi}$ is explicitly given by \eqref{fdiff}, which is the limit of \eqref{phiphi} at $r\rightarrow \infty$. Variable $\tilde{\varphi}$ is just the conventional angular variable of the BTZ geometry, while the radial variable $\tilde{r}$ emerge as an integration constant parametrizing the Killing vector trajectory \eqref{Kxi}.
At leading order in  $1/r$ expansion we have 
\bea
dr=-f' r d\tilde{\varphi},\quad d\varphi =f d\tilde{\varphi}\qquad  \Rightarrow\qquad  \tilde{r}= rf.
\eea

If $f$ is not sign-definite, corresponding $u(\varphi)$ belongs to a special class of co-adjoint orbits, which, upon quantization, corresponds to a reduced representation of Virasoro algebra. As the dual CFT$_2$ with $c\gg 1$ has no such representations  in its spectrum (except for the vacuum module), we disregard the possibility of $f$ vanishing at some $\varphi$. We revisit this question later in section \ref{sec:other}.

Finally, we discuss the  condition for the geometry \eqref{met1} to be non-singular. For that we must require that the black hole horizon ``hides'' the singularity $g_{\varphi\varphi}=0$, if it exists, as well as the point $r=0$ where $g_{rr}$ diverges. This leads to the following two conditions
\bea
r_h^2\geq -{\ell^2 u\over 4},\qquad r_h^2\geq 0,
\eea
which can be rewritten as 
\bea
u_0 f_0^2\geq f'^2- f f'',\qquad u_0 f_0^2\geq f'^2.
\eea
Since $u_0 f_0^2$ is a constant, the inequality implies that $u_0 f_0^2$ is larger or equal than the maximum value of $f'^2- f f''$ and $f'^2$. The maximum of $f'^2$ occurs at $f''=0$, hence the maximum of $f'^2- f f''$ is not smaller than the maximum of $f'^2$. We therefore conclude that the first inequality implies the second and the remaining constraint is
\bea
u_0 f_0^2\geq \max_\varphi f'^2- f f''. \label{smoothnesscondition}
\eea
From here follows that $u_0\geq 0$, which is consistent with the expectation that smoothness of the  conventional BTZ geometry requires $u=u_0\geq 0$. 
Positivity of $u_0$ is necessary, but not sufficient.  We will show in section \ref{sec:Z} that there are one-cut solutions for which $u_0>0$ yet  \eqref{smoothnesscondition} is not satisfied. 

\subsection{Black hole thermodynamics}

The black hole horizon area is given by the integral of the induced metric on the horizon, $r^2=r_h^2(\varphi)$,
\bea
S={1\over 4G_N}\int_0^{2\pi}d\varphi\sqrt{h_{\varphi\varphi}}={\ell\over 4G_N}\int_0^{2\pi} {d\varphi \over f}{u f-f''\over u f-2 f''} (u f^2+f'^2-2f f'')^{1/2},
\eea
where Newton's constant $G_N$ is related to the AdS radius $\ell$ and dual theory central charge $c$ as follows, $\ell/G_N={2\over 3}c$ \cite{Brown:1986nw}.
After using \eqref{u0} and noticing that ${f''\over u f^2-2 f f''}$ is a full derivative $(\arctan(f'/\sqrt{u_0 f_0^2}))'/\sqrt{u_0 f_0^2}$
we find assuming $f_0>0$,
\bea
\label{S}
S={\pi c\over 3}\sqrt{u_0}. \label{S}
\eea
Thus entropy is constant along the co-adjoint orbit, which is expected -- the horizon area is invariant under diffeomorphisms and is given by \eqref{S} in the BTZ case.  The   Using identification  \eqref{spectrum}  we see that it agrees with the Cardy formula. 
To obtain the temperature we analytically continue the solution to imaginary time, $t\rightarrow \tau= i t$ and require the absence of conical singularity at the horizon. This imposes periodicity $\tau\sim \tau+T^{-1}$, where 
\bea
\label{T}
(2\pi T)^2=u f^2+f'^2-2 f f''=u_0 f_0^2.
\eea
We will show now that in full generality the KdV-charged black holes satisfy the thermodynamic identity
\bea
TdS=d{\mathcal H}_G \label{thermo}, \qquad {\mathcal H}_G=\frac{c}{12}{\mathcal H},
\eea 
where the differential is with respect to an arbitrary small variation of $u$. Because  \eqref{thermo} is linear in $\delta u$, we can split the variation of $u$ into two parts, along the co-adjoint orbit, and in any transversal direction. Since the Poisson bracket is non-degenerate on the co-adjoint orbit and $\{{\mathcal H},u\}=0$, the RHS of \eqref{thermo} with respect to any variation of $u$ along the orbit will vanish. This is consistent with the fact that the value of entropy \eqref{S} is constant along the orbit, and hence $dS=0$. Now we need to consider a transversal direction. We would like to parametrize $u(\varphi)$ using $u_0$ and $f$ as in \eqref{u0}, and vary $u_0$, which will correspond to moving between different orbits. In this case
\bea
\label{du0}
\delta u =d u_0 {f_0^2\over f^2}, 
\eea
and 
\bea
\label{dH}
d{\mathcal H}=\int_0^{2\pi} d\varphi {\delta \mathcal H\over \delta u}\delta u=d u_0 f_0,
\eea
where we used \eqref{fu}. This is in agreement with $TdS/du_0=\frac{c}{12}f_0$, which follows from \eqref{S} and \eqref{T}. 

In the discussion above we have implicitly used that $f_0$, or equivalently $f$, is positive. What happens when $f$ is sign-definite but negative? Since the entropy and the temperature are defined geometrically, as the horizon  area and periodicity of $\tau$ coordinate, these quantities are positive-definite and thus given by \eqref{S} and \eqref{T}. At the same time the variation  \eqref{dH} with respect to \eqref{du0} will be negative, and therefore instead of \eqref{thermo} one would have $TdS=-d{\mathcal H}_G$. This is a violation of the first law of thermodynamics, but it does not mean the corresponding geometry is pathological. Indeed the black hole geometry  (\ref{met1}, \ref{met2}) does not depend on $f$, but only the equivalence class $\pm f$. Hence the same Lorentzian signature black hole can be understood as a solution (state) in different theories with different Hamiltonians ${\mathcal H}$. In certain cases  when ${\delta {\mathcal H}\over \delta u}$ is positive,   thermodynamic identity will be satisfied, whenever  ${\delta {\mathcal H}\over \delta u}$ is negative it will be violated. In other words, the same geometry may satisfy or violate the first law, depending on the choice of ${\mathcal H}$. This clearly shows that as the Lorentzian signature geometries black holes with negative $f$ are non-pathological, provided the regularity condition \eqref{smoothnesscondition} is satisfied. 

The problem with  negative $f$ is the problem of the Euclidean geometry interpretation. Normally we interpret an Euclidean geometry as a saddle point of the Euclidean path integral, which would require  $TdS=d{\mathcal H}_G$. To maintain this interpretation we propose that whenever $f<0$, correct interpretation would be to assign temperature negative value, such that in full generality
\bea
T={\sqrt{u_0}f_0\over 2\pi}.
\eea
Then the thermodynamic identity is restored, while corresponding Euclidean black hole is a saddle point configuration of the path integral evaluating ${\rm Tr}(e^{-{\mathcal H}_G/T})$ on the gravity side. Last formula should be understood in a formal sense because, provided ${\mathcal H}$ is bounded from below, the sum ${\rm Tr}(e^{-{\mathcal H}_G/T})$ is not convergent for negative $T$. 

To summarize, while black hole configurations with negative $f$ and  temperature are well-defined in Lorentzian signature, we do not include their contribution toward Euclidean path integral, as is the case for one-cut geometries discussed in section \ref{sec:Z}.

\subsection{Other solutions} 
\label{sec:other}
If $f$ is sign-definite, there is always the diffeomorphism \eqref{fdiff} which brings $u$ to a constant form, and the geometry becomes the BTZ black hole or (thermal) AdS$_3$. But there are also solutions when $f$ vanishes at certain points.  For example $u=-1$, $f=\cos(\varphi)$ results in the following metric
\begin{align}
    ds^2=-\rho^2 \cos^2\varphi\, dt^2+\frac{d\rho^2}{\rho^2+\ell^2}+\rho^2 d\varphi^2,\qquad \rho=r-\frac{\ell^2}{4r},
\end{align}
with the boundary, upon the continuation to Euclidean signature, being a torus degenerated into two spheres, $ds^2=\cos^2\varphi\, dt^2+d\varphi^2$, $0\leq \varphi\leq 2\pi$. Since the boundary geometry is not a torus, we omit such configurations from the Euclidean path integral associated with $\mathcal Z$. In fact if $f$ vanishes for some $\varphi$, either $f''$ also vanishes, in which case boundary geometry becomes non-trivial (in a sense that it is not a torus), or, when $f''\neq 0$, there is no horizon but a singularity at $r=0$. To conclude, while the geometries with non sign-definite $f$ might be interesting in their own right, we do not expect them to contribute in the calculation of $\mathcal Z$.

Another interesting possibility is the non-static configurations, i.e.~some time-dependent solutions of the higher KdV equation \eqref{EOM}. For such a solution to be a saddle point of the gravity Euclidean path integral, upon analytic continuation $t\rightarrow \tau=it$, it should be defined on a torus, i.e.~be double-periodic with respect to  $\varphi\sim \varphi+2\pi$ and $\tau\sim \tau+1/T$.
In the original CFT with ${\mathcal H}=\hat{Q}_1$ this is not possible because \eqref{EOM} reduces to the analyticity condition
\bea
\label{holomorphic}
(\partial_\varphi-i \partial_\tau)u=0,
\eea
and there is no non-singular analytic functions on a torus except for a constant. Hence BTZs are the only configurations contributing to the Euclidean path integral. Similarly, $u$ can not be  one-cut solution even if the Hamiltonian is deformed by KdV charges, because in that case action of any $Q_{2k+1}$ is proportional to $\partial_\varphi u$. Hence for any ${\mathcal H}$ we again end up with \eqref{holomorphic} after an appropriate rescaling of $\tau$. There is nevertheless a hypothetical possibility that a more complicated multi-cut solutions, upon analytic continuation $t\rightarrow \tau=it$ would be double-periodic and non-singular for any $\varphi,\tau$. Then, provided it can be smoothly extended into the bulk,  such a solution would be a new non-trivial saddle point configuration of the  Euclidean gravity path integral. 

\section{CFT interpretation and thermalization}
\label{sec:CFT}
We start with the AdS$_3$ geometry $u=-1$ and its analytic continuation to Euclidean signature, thermal AdS space. As a Lorentzian-signature geometry it is dual to the vacuum state $|0\rangle$ in the original CFT, as follows from  \eqref{spectrum}.
This state is not necessarily the ground state in the theory with a deformed Hamiltonian. Accordingly, the Euclidean solution is a saddle point of the Euclidean gravity path integral, but is usually not  the leading one. 

The Lorentzian-signature black hole geometries discussed in the previous section are holographically dual to KdV GGE states \eqref{GGE} in field theory, but this identification requires additional clarification. First, the state  \eqref{GGE} is stationary in a theory with any Hamiltonian, would it be $H=L_0-c/24$, $H \propto \sum_k \tilde{\mu}_k \hat{Q}_{2k+1}$, or any other linear combination of qKdV charges. On the gravity side a geometry associated with  $u(\varphi)$ is a static  solution of \eqref{EOM} when the Hamiltonian is ${\mathcal H}$. As a geometry in a theory of gravity  with the time evolution generated by some other  linear combination of $Q_{2k+1}$ it is a time-dependent soliton solution. As was discussed in section \eqref{sec:Novikov}, the trajectory generated by a general linear combination of $Q_{2k+1}$ would densely cover the Jacobian of the solutions with the same values of $Q_{2k+1}$ as the original $u$, such that time averaged value of any observable would be equal to the average over the Jacobian. (For mathematical rigor, this requires $u$ to be a finite-zone solution, otherwise the ``Jacobian'' would be infinite-dimensional.) This is a classical counterpart of 2d CFT generalized Eigenstate Thermalization Hypothesis established in \cite{GETH}, which states that expectation value of an observable from the vacuum family, i.e.~made of $T$, in a qKdV eigenstate is completely specified by the values of the qKdV charges associated with this eigenstate. Thus, the GGE state  \eqref{GGE} is dual not to a particular geometry, but to a probabilistic average of all possible geometries built with $u(\varphi)$ which correspond to the same spectral curve, i.e.~have the same values of $Q_{2k+1}$.

There is a qualitative difference between $Q_1$ and all other $Q_{2k+1}$, $k\geq 1$. While all higher KdV charges in a general case  generate an ergodic flow on the Jacobian leading to eventual thermalization at the   classical level, the flow generated by $Q_1$ is always $2\pi$-periodic, and thus non-ergodic. The counterpart of that at the quantum level is that the spectrum of $\hat{Q}_1=L_0-c/24$ is highly degenerate, while any other $\hat{Q}_{2k+1}$ removes this degeneracy. Thus, the diagonal part of generalized ETH assures {\it eventual} thermalization for any initial state, provided the CFT Hamiltonian includes higher qKdV charges, but not if $H=\hat{Q}_1$.\footnote{Our definition of thermalization  is  based on time-averaging of local observables in CFT. Thus different initial configurations $u(\varphi)$, even if they belong to the same co-adjoint orbit, lead to different time-averaged values. This is different from the approach  of  \cite{Banerjee:2018tut,Vos:2018vwv} which  define ``thermalization'' by averaging over all possible observables on the AdS boundary, such that all $u(\varphi)$ belonging to the same orbit ``thermalize'' to the same final state.}

Among the finite-zone solutions there are also trivial solutions $u=u_0={\rm const}$. The corresponding geometry is the conventional BTZ black hole, albeit understood as a state in a theory with the deformed Hamiltonian. These black holes were initially considered in \cite{perez2016boundary} and more recently revisited in \cite{Erices:2019onl}. Since a constant solution is invariant under the action of all $Q_{2k+1}$ we readily identify corresponding Lorentzian geometry as a holographic dual of a qKdV eigenstate, or, more accurately, exponentially many eigenstates with an approximately equal energy, where the log of the number of states is given by \eqref{S}. These states on the CFT side are discussed in detail in \cite{maloney2018generalized,GGE,GGE2}. The CFT calculation of $\mathcal Z$ outlined there can be easily seen to mirror the contribution of the BTZ geometries to the Euclidean gravity path integral. 

To complete the holographic dictionary we need to know how to map chemical potentials $\tilde{\mu}_{2k+1}$ of the GGE state to a particular black hole geometry, or, more accurately, probabilistic superposition of all geometries with $u(\varphi)$ being associated with the same spectral curve. If $u$ is a solution of \eqref{EOM} with some particular Hamiltonian \eqref{AdSH}, it is not necessarily true that the chemical potentials $\tilde{\mu}_{2k+1}$ of a dual GGE  are given  by \eqref{H}. This is already clear from the fact that the same $u(\varphi)$ is a solution of \eqref{EOM} for infinitely many different $\mathcal H$. Rather a correct CFT dual is the GGE state with the chemical potentials $\tilde{\mu}_{2k+1}$ chosen such that the expectation values of $\hat{Q}_{2k+1}$,  
\bea
{\rm Tr}(\rho_{\rm GGE}\, \hat{Q}_{2k+1})=-\partial_{\tilde{\mu}_{2k+1}}\ln \mathcal Z,
\eea 
match those of its gravity dual counterpart. In the large $c$ limit one may assume that $\mathcal Z$ is given by its leading saddle, a configuration which minimizes free energy ${\mathcal F}={\mathcal H}_G-S$ (we take temperature $T$ to be one because it always can be absorbed into $\mu_{2k+1}$). A non-trivial question then would be to find $\mu_{2k+1}$ such that given $u(\varphi)$ is the leading saddle where minimum of ${\mathcal F}$ is achieved. 

\section{KdV-generalized partition function with $\mu_1,\mu_3\neq 0$}
\label{sec:Z}
In this section we  calculate KdV-generalized partition function $\mathcal Z(\mu_1,\mu_3)$ on the gravity side, assuming the only contributing configurations are those discussed in section \ref{sec:BH}. The Hamiltonian $\mathcal H=\mu_1 Q_1+\mu_3 Q_3$ for $\mu_3>0$ and arbitrary $\mu_1$ is bounded from below, hence $\mathcal Z$ is well-defined. We fix temperature to be $T=1$  since it can be always absorbed into $\mu_i$.

First contribution comes from the thermal AdS configuration $u=-1$, which has free energy 
\bea
{\mathcal F}_{\rm AdS}={\mathcal H}_G={c\over 12}(\mu_3-\mu_1). \label{FAdS}
\eea  
This calculation parallels the CFT calculation, where the value of $\hat{Q}_3$ evaluated in any primary state is given by $\hat{Q}_1^2-\hat{Q}_1/6+c/1440$. Thus, evaluated in vacuum, $\langle0|\hat{Q}_3|0\rangle\sim (c/24)^2+O(c)$. Going back to \eqref{FAdS}, unless $\mu_1$ is large, free energy is positive, and therefore there is no Hawking-Page transition. Here we disagree with \cite{perez2016boundary,Erices:2019onl}, where thermal AdS  free energy was assigned negative sign, ${\mathcal F}_{\rm AdS}=-{c\over 12}\mu_3$ for $\mu_1=0$, and regard that as a mistake. 

Next, we are looking for the BTZ geometry $u=u_0\geq 0$ with temperature $T=1$, 
\bea
\label{Tu0}
T={\mu_1+2\mu_3 u_0\over 2\pi}\sqrt{u_0}=1.
\eea
This equation always has a unique solution $u_0(\mu_3,\mu_1)$ for $\mu_3>0$ and arbitrary $\mu_1$. Free energy in this case is given by 
\bea
{\mathcal F}_{\rm BTZ}={c\over 12}\left(\mu_3 u_0^2+\mu_1 u_0-4\pi \sqrt{u_0}\right)=-{c\over 12}(3\mu_3 u_0^2+\mu_1 u_0).
\eea

As we will see now there are no non-trivial  black hole solutions with $u(\varphi)\neq {\rm const}$ contributing in this case. This conclusion is intuitive because $\mathcal H$ includes no derivatives, hence the saddle point is $u={\rm const}$, but the precise argument is much more elaborate. This is partially because entropy depends on $u$ in a non-trivial and non-local way. The static solution $\{{\mathcal H},u\}=0$ is necessarily a one-zone potential, with the stabilizer vector field 
\bea
\label{ff}
f=2\mu_3 u+\mu_1. 
\eea
For the geometry to be smooth and temperature positive
we must require $f$ to be sign-definite and positive. Combining \eqref{constraint} together with \eqref{ff},  positivity of $f$ yields the following constraint
\bea
\forall \varphi,\quad u-s_1=u(\varphi)-u_1-u_2-u_3> 0, \label{c1}
\eea
where $u_i$ are the parameters of the one-zone solution introduced in section \ref{sec:1zone}. The non-constant solution oscillates between $u_1$ and $u_2$, hence \eqref{c1} implies
\bea
\lambda_0<0,
\eea
where $\lambda_0$ is introduced in \eqref{lambdau}.  This contradicts the smoothness condition \eqref{smoothnesscondition}, which requires that the constant representative of the orbit $u_0$ to which $u(\varphi)$ belongs must be positive. As follows from \eqref{h} for that point $\lambda=0$ must be on the branch-cut of $p(\lambda)$, which goes from minus infinity to $\lambda_0$. Hence smoothness of the bulk geometry requires $\lambda_0>0$. 

Absence of saddle points with non-constant $u(\varphi)$ when only $\mu_1,\mu_3\neq 0$, and emergence of  saddle points characterized by $Q_{2k+1}\neq Q_1^k$ when $\mu_5\neq 0$ discussed in the next section is in agreement with the analysis of  \cite{GETH} of small fluctuations around the BTZ background.

\begin{figure}
	\includegraphics[width=\textwidth]{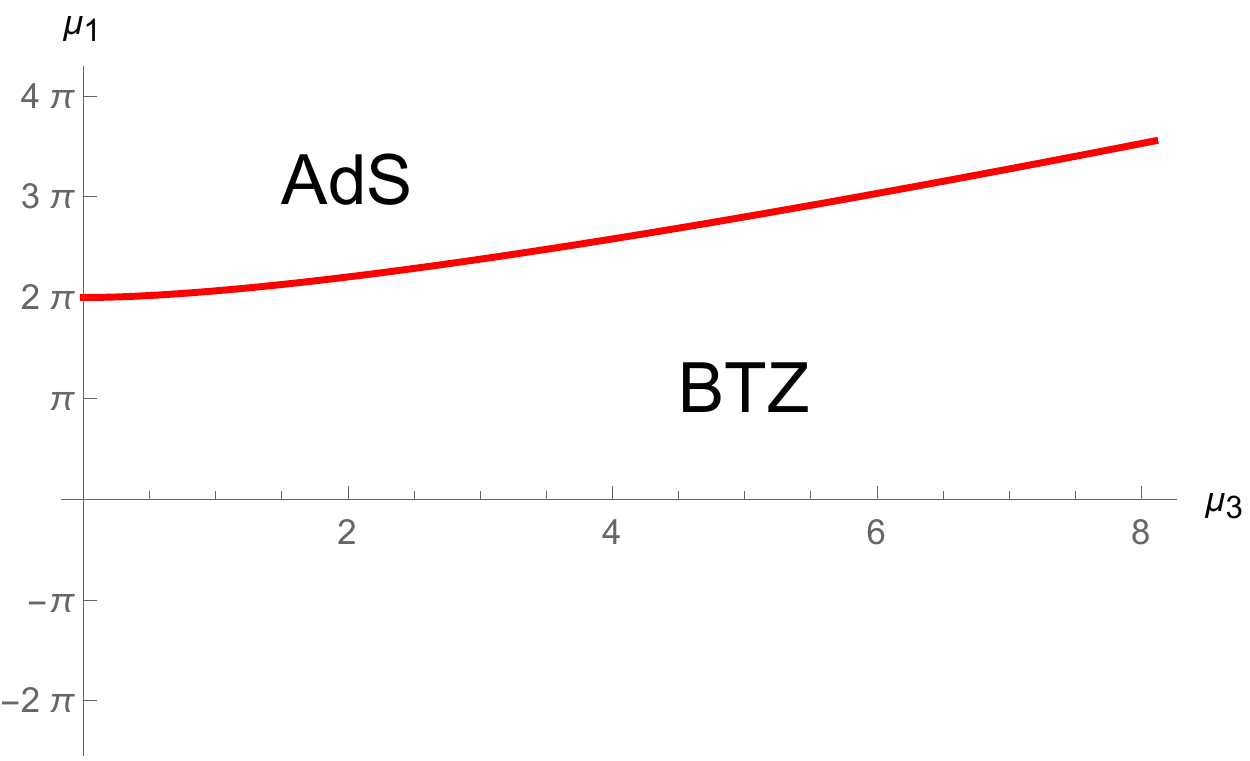}
	\caption{
		\label{mu1-mu3}
		The phase-diagram on $\mu_3$-$\mu_1$ plane with $\mu_3\geq 0$. (Euclidean) BTZ is the leading contribution to the generalized partition function \eqref{Z13} in the region below the red curve. In particular, BTZ is always the leading saddle for non-positive $\mu_1$. 
		}
\end{figure}

Thus, thermal AdS and the BTZ geometries are the only contributions when only $\mu_1,\mu_3\neq 0$, yielding in the infinite $c$ limit
\bea
\label{Z13}
{\mathcal Z}=e^{-{\mathcal F}_{\rm AdS}}+e^{-{\mathcal F}_{\rm BTZ}}.
\eea
Finally we investigate the possibility of Hawking-Page transition when the contributions become equal. Equating ${\mathcal F}_{\rm AdS}={\mathcal F}_{\rm BTZ}$, together with \eqref{Tu0}, gives an algebraic equation for $\mu_1$ in terms of $\mu_3$, which divides the $\mu_3$-$\mu_1$ plane into two regions,  with the leading contribution marked in Fig.~\ref{mu1-mu3}.

It is important to point out that the condition $f>0$ is necessary only to ensure positivity of temperature $T>0$. As a geometry in Lorentzian signature  (\ref{met1},\ref{met2}) is well-defined and smooth also for negative $f$ as long as the condition \eqref{smoothnesscondition} is satisfied. For one-cut $u(\varphi)$ and $f$ given by \eqref{ff}  the condition \eqref{smoothnesscondition} can be satisfied  provided $f$ and $T$ are negative. Hence we arrive at an unusual situation when ${\mathcal H}=\mu_1 Q_1+\mu_3 Q_3$ may have a static solitonic solution which nevertheless does not correspond to a saddle point configuration upon analytic continuation to Euclidean signature. This is because the Euclidean geometry is a saddle point for a theory with the Hamiltonian $-{\mathcal H}$, which, in our case, would be unbounded from below.

\section{Dominance of multi-cut solutions}
\label{sec:5}
The logic of this subsection is opposite to the previous one. We start with a particular one-cut solution $u(\varphi)$ characterized by some fixed $k,s_1, q$, and make sure that corresponding geometry are smooth. We also find a Hamiltonian of the form ${\mathcal H}=\mu_1 Q_1+\mu_3 Q_3+\mu_5 Q_5$, $\mu_5>0$, such that the black hole geometry build with this $u(\varphi)$  has a smaller free energy than any $u={\rm constant}$, i.e.~thermal AdS or BTZ configuration.
 
We first find all ${\mathcal H}=\mu_1 Q_1+\mu_3 Q_3+\mu_5 Q_5$ such that given $u(\varphi)$ is a static solution.
For a solution of \eqref{potential} the stabilizer vector field $f$ such that $\D f=0$ is always proportional to $f\propto u-s_1$. 
Any $Q_{2k+1}$ acting on a one-cut solution generates a flow proportional to $\partial_\varphi$, for example
\bea
{c\over 24}\{Q_5,u\}=(2s_1^2-s_2)u'.
\eea
Here and below $u_i$ and $s_2,s_3$  are understood as functions of $k,s_1, q$, given by (\ref{def:u1}-\ref{def:s3}).
Therefore $\{{\mathcal H},u\}=0$ for ${\mathcal H}=\mu_1 Q_1 + \mu_3 Q_3+\mu_5 Q_5$ as long as
\bea
\label{m1}
\mu_1+2 s_1 \mu_3+(2s_1^2-s_2)\mu_5=0.
\eea
Equation \eqref{m1} can be used to express $\mu_1$ in terms of $\mu_3,\mu_5$. An overall rescaling of $\mu_i$ must be fixed by the requirement $T=1$.
From the definition $f=2\pi {\delta H\over \delta u}$ we find
\bea
\label{f_1cut}
f=2(\mu_3+\mu_5\, s_1)(u-s_1).
\eea
As discussed in the previous subsection, since $u$ oscillates between $u_1$ and $u_2$, the combination $u-s_1$ should be negative in order to have $\lambda_0>0$. For $T>0$ we therefore must require $\mu_3+\mu_5\, s_1<0$ such that $f$ is positive. 
Using \eqref{T} for one-cut solution we readily find
\bea
\label{fixm3}
T=-(\mu_5 s_1 +\mu_3){\sqrt{2(s_1 s_2-s_3)}\over 2\pi}=1.
\eea
This fixes $\mu_3$ in terms of $\mu_5$.  
Hence for any given one-cut solution specified by $k,s_1,q$ we have a family of Hamiltonians parametrized by $\mu_5$ such that $\{{\mathcal H},u\}=0$ and $T=1$, while 
$\mu_1,\mu_3$ is fixed in terms of $\mu_5$ by \eqref{m1}  and \eqref{fixm3}.

Now we consider the  smoothness condition \eqref{smoothnesscondition},  which is independent of the overall rescaling of $f$ and can be rewritten as $u f^2- ff''>0$. Taking into account that $u-s_1$ must be negative, this yields
\bea
(u-s_1)^2-s_1^2-s_2<0.
\eea
Since $u_1\leq u\leq u_2$ and $u-s_1<0$, in the condition above we can take $u=u_2$, thus reducing it to a quadratic inequality on $s_1/k^2$. Together with the linear inequality $u_1+u_2>0$, which is equivalent to $u_0>0$,  this becomes 
\bea
\label{cond1-s1}
\left\{
\begin{array}{l}
 4s_1^2-4k^2\left(\theta_2(0;q)^4+\theta_3(0;q)^4\right)s_1-k^4(7\theta_2(0;q)^8+7\theta_3(0;q)^8+\theta_4(0;q)^8)>0 \\
s_1>k^2\left(\theta_3(0;q)^4+\theta_4(0;q)^4\right)\geq 2k^2.
\end{array} 
\right.
\eea 
The allowed region of $s_1/k^2$ as a function of $q$ is shown in Fig.~\ref{allowpara}. An additional analysis shows that in this region $\mu_1,\mu_5>0$ and $\mu_3< 0$, see Appendix \ref{App}.
\begin{figure}
	\begin{center}
	\includegraphics[scale=0.8]{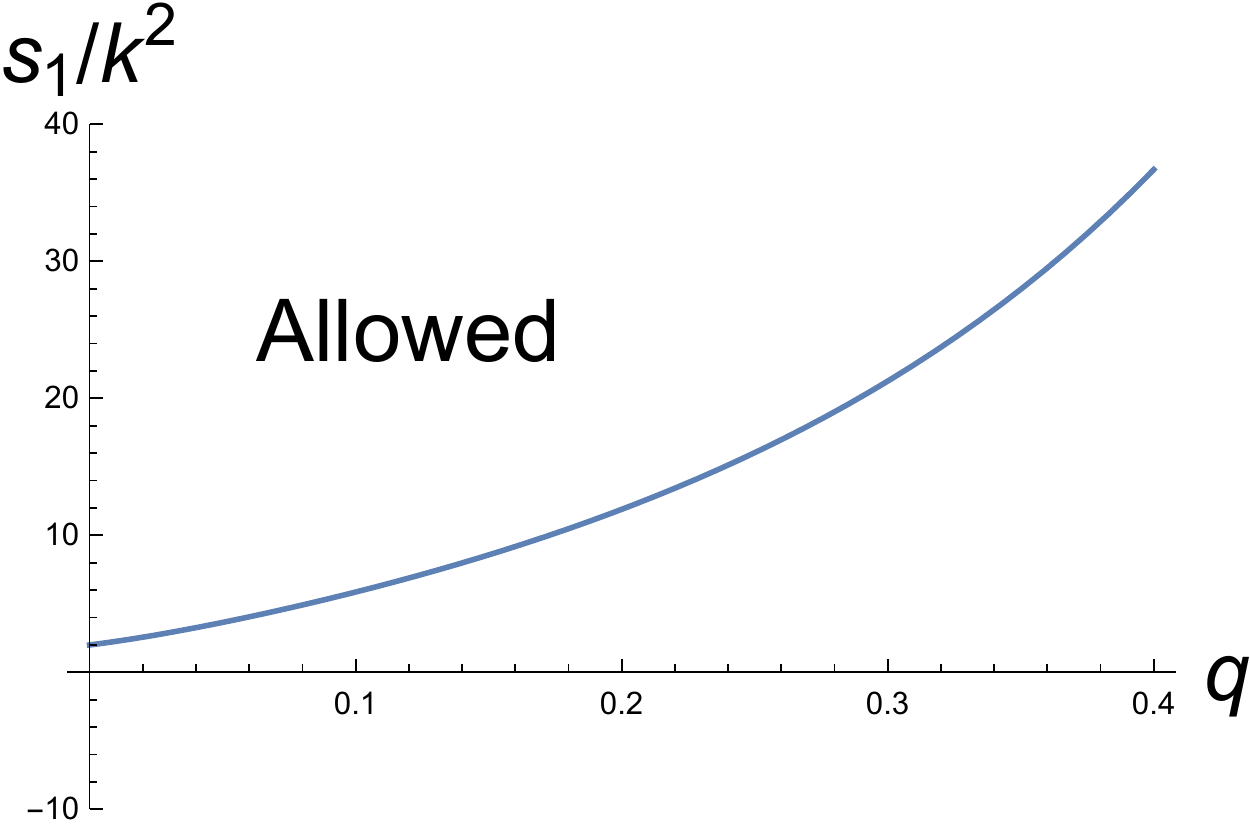}
	\caption{The allowed region of $s_1/k^2$ and $q$ satisfying both \eqref{cond1-s1}. 
	The region above the blue curve is allowed. Thus, at least, $s_1$ should be greater than $2k^2$.
    \label{allowpara}}
    \end{center}
\end{figure}

It is easy to see that second condition in  \eqref{cond1-s1}, which is $u_0>0$, is not implying  first condition, which is a consequence of \eqref{smoothnesscondition}. Hence, positivity of $u_0$ is necessary but not sufficient for the geometry to be regular.

Free energy of one-cut solution is given by 
\begin{align}
\mathcal{F}_\text{1-cut}(\mu_5)=\frac{c}{12}(\mu_1 Q_1 +\mu_3 Q_3 +\mu_5 Q_5)-\frac{\pi c}{3}\sqrt{u_0},
\end{align}
where $u_0$ is given by \eqref{u0} and $f$  is given by \eqref{f_1cut}.
The value of the  charges $Q_1, Q_3, Q_5$ for the one-cut solution are given in \eqref{Q1_onecut} and \eqref{Q3Q5_onecut}. 
Thus, for a given one-cut solution with  fixed $k, s_1, q$, free energy $\mathcal{F}_\text{1-cut}$ is a function of $\mu_5$, while $\mu_1$ and $\mu_3$ are fixed by \eqref{m1} and \eqref{fixm3}. 

We want to compare  free energy $\mathcal{F}_\text{1-cut}$ with those of thermal AdS, and BTZ solutions corresponding to ${\mathcal H}(\mu_i)$ with the same values of chemical potentials and $T=1$. Free energy of thermal AdS is given by
\begin{align}
\mathcal{F}_\mathrm{AdS}(\mu_5)&=\frac{c}{12}(-\mu_1 +\mu_3 -\mu_5)
=-\frac{c}{12}\left((s_1+s_2+1)\mu_5+\frac{2\pi(2 s_1+1)}{\sqrt{2(s_1 s_2-s_3)}}\right).
\end{align}

To BTZ solutions $u=u_0$ with temperature $T=1$  must satisfy
\begin{align}
1=T=\frac{\mu_1+2\mu_3 u_0 +3\mu_5 u_0^2}{2\pi}\sqrt{u_0}.
\end{align}
Since $\mu_5>0, \mu_3<0, \mu_1>0$,  there are at most three solutions. We choose $u_0$ such that the free energy 
\begin{align}
\mathcal{F}_\mathrm{BTZ}(\mu_5)=\frac{c}{12}(\mu_1 u_0 +\mu_3 u_0^2 +\mu_5 u_0^3)-\frac{\pi c}{3}\sqrt{u_0}
\end{align}
is smallest. 

\begin{figure}
	\begin{center}
		\includegraphics[scale=0.8]{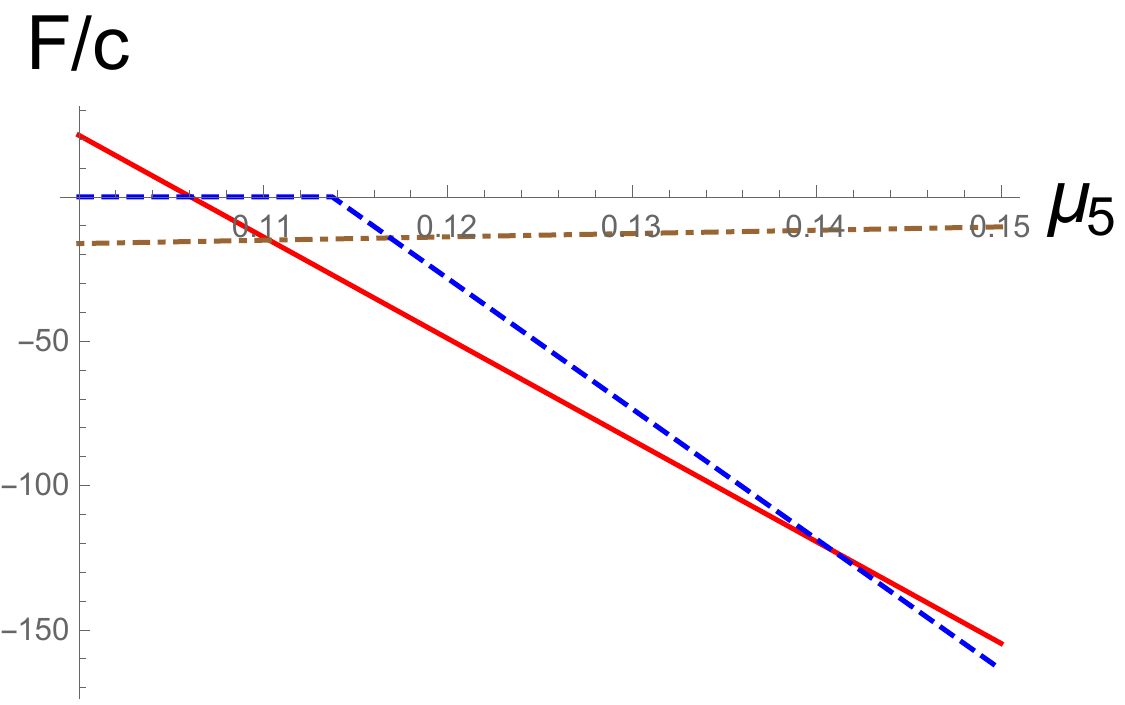}
		\caption{Free energy of the one-cut, ``smallest'' BTZ, and thermal AdS configurations as a function of $\mu_5$ for $k=1, s_1=38, q=0.6$. The solid red, dashed blue and dotted brown lines represent $\mathcal{F}_\text{1-cut}/c$,  $\min \mathcal{F}_\mathrm{BTZ}/c$, and $\mathcal{F}_\mathrm{AdS}/c$ respectively. There is a region of $\mu_5$ where  free energy $\mathcal{F}_\text{1-cut}$ (solid red line) is smaller than the two others.
		\label{fig:free}}
	\end{center}
\end{figure}

We plot free energies for one-cut, thermal AdS, and ``smallest'' BTZ solutions  for some $k, s_1, q$ as a function of $\mu_5$ in Fig.~\ref{fig:free}.  
There is clearly  a region in the parameter space such that the one-cut solution has the smallest free energy among the considered configurations. Whether it is the leading saddle in this case remains unclear because the Hamiltonian in question  also has two-cuts static solutions.  

\section{Discussion}
\label{sec:Discussion}

In this paper we have constructed black hole geometries in AdS$_3$ which carry charges under the KdV symmetries. These geometries are static solutions in the  theory of pure gravity  in AdS$_3$ with the deformed boundary conditions, such that the Hamiltonian in the dual CFT is a linear combination of qKdV charges. Each geometry is specified by $u(\varphi)$ which is a static solution of a higher KdV equation, i.e.~a finite zone  Novikov solution when there are only finite number of KdV charges involved. Accordingly many properties of the black holes, including the first law of thermodynamics, follow directly from the geometry of the co-adjoint orbit of Virasoro algebra and basic properties of the KdV equations. Nevertheless there are several key ingredients which come from the bulk and are external to  the classical KdV theory. First, this is  the Bekenstein-Hawking entropy  \eqref{S}. 
Second, smoothness of the geometry in the bulk yields the regularity condition \eqref{smoothnesscondition}, which goes beyond the condition expected on the grounds that the new geometries are diffeomorphic to BTZ ones, namely that $u(\varphi)$ belongs to the Virasoro co-adjoint orbit ${\rm diff}\, \Sphere^1/\Sphere^1$ with $u_0>0$. It would be interesting to understand what this new condition means in terms of the finite zone solutions. Finally, thermodynamic identity requires $T>0$ and hence $f>0$. While this is not a condition on $u(\varphi)$, this is a condition on the pair ${\mathcal H}$ and $u$.
More generally, the main questions formulated in this paper, of identifying $u$ which is the leading saddle for a given ${\mathcal H}$, and the question of identifying ${\mathcal H}$ such that given $u$ is its leading saddle, are the well posed new questions in the context of KdV theory. 

The holographic dual  of the classical GGE state discussed in this paper is the  KdV analog of the GGE for another classical integrable model  recently discussed in \cite{spohn2019generalized,bulchandani1905gge}. The next logical step here would be to develop a theory of generalized hydrodynamics describing long-wave dynamics of states locally deviating from the GGE \cite{castro2016emergent,ilievski2017ballistic,piroli2017transport,bastianello2018generalized,doyon2019generalized}. This description should be valid both in the classical limit of field theory, and in the bulk, where it would describe the dynamics near a black hole background. In this context it would be interesting  to see if the entropy \eqref{S} can be given a microscopic interpretation in terms of the classical field theory of $u(\varphi,t)$, providing geometric interpretation to black hole microstates. 

The Euclidean black hole geometries are the classical saddles of the ``generalized'' Alekseev--Shatashvili  path integral \cite{alekseev1989path} decorated by higher KdV charges. A natural question would be to quantize small fluctuations around the black hole background to obtain $1/c$ corrections. This was done in \cite{GGE2} on the filed theory side for the conventional BTZ background, but the case of a non-constant $u(\varphi)$ needs to be treated separately. On a different note,  in the limit of large temperature the boundary torus will reduce to a circle, while the Alekseev--Shatashvili path integral should reduce to the Schwarzian theory of SYK/JT gravity \cite{maldacena2016remarks,stanford2017fermionic,callebaut2019entanglement}. It would be interesting to see if the geometries discussed in this paper would give rise to non-trivial saddle point configurations of the generalized Schwarzian theory which includes higher order operators, in particular $T\bar{T}$ deformation.

\appendix
\section{Appendix: Signs of $\mu_1$, $\mu_3$ in the allowed region} 
\label{App}
In this appendix we show that $\mu_1>0$ and $\mu_3<0$ in the region specified by the smoothness condition \eqref{cond1-s1}.
First, $\mu_3< 0$ which simply follows from $\mu_3+\mu_5 s_1<0$, see \eqref{fixm3}, and $s_1>2k^2$ imposed by \eqref{cond1-s1}. 

\begin{figure}
	\begin{center}
		\includegraphics[scale=0.7]{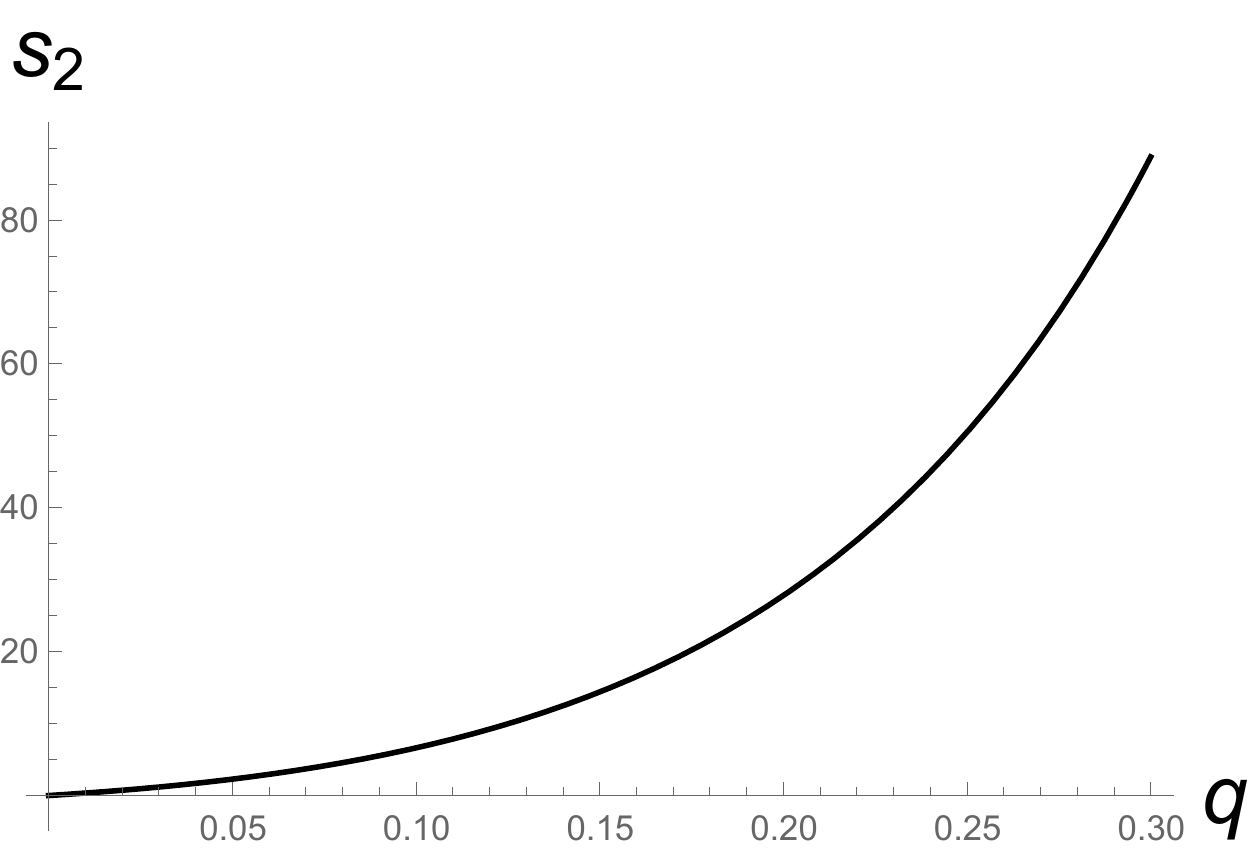}
		\caption{Plot of $s_2$ as a function of $q$ for $s_1(q)$ which saturates the first constraint of \eqref{cond1-s1}, i.e.~along the blue curve in Fig.~\ref{allowpara}. 
			For all $q>0$, $s_2$ is positive, and $s_2=0$ only for $q=0$. It means that $s_2\geq 0$ in the allowed region in Fig.~\ref{allowpara}. 
		\label{fig:s2}}
	\end{center}
\end{figure}

Next, using \eqref{m1} we find
\begin{align}
\mu_1=\mu_5 s_2-2 s_1(\mu_3+\mu_5 s_1 ).
\end{align}
Since $\mu_5>0$ and $\mu_3+\mu_5 s_1<0$, we find $\mu_1>0$ if $s_2\geq 0$. Note that $s_2$ is given by \eqref{def:s2} and it is a monotonically increasing function of $s_1>0$ for a fixed $q$. We check that $s_2$ as a function of $q$ for $s_1$ saturating the first constraint of \eqref{cond1-s1} is positive as shown in Fig~\ref{fig:s2}. Thus, $s_2\geq 0$ in the entire region where the inequalities \eqref{cond1-s1} are satisfied. We therefore have $\mu_1> 0$. 

\acknowledgments
We thank Daniel Jafferis, Nikita Nekrasov, Albert Schwarz, and Xi Yin  for discussions. 
AD is supported by the National Science Foundation under Grant No. PHY-1720374.  AD is grateful to KITP for hospitality, where this work has been completed. The research at KITP was supported in part by the National Science Foundation under Grant No.~NSF PHY-1748958. SS acknowledges the hospitality of Yukawa Institute for Theoretical Physics, where a part of this work was done and was presented. SS also thanks members of Particle Physics Theory Group at Osaka Univ.~for an opportunity to present this work and for discussions.


\bibliographystyle{JHEP}
\bibliography{KdVBH}

\providecommand{\href}[2]{#2}\begingroup\raggedright\begin{thebibliography}{10}

\bibitem{calabrese2006time}
P.~Calabrese and J.~Cardy, \emph{Time dependence of correlation functions
  following a quantum quench}, {\emph{Physical review letters} {\bfseries 96}
  (2006) 136801}.

\bibitem{calabrese2007quantum}
P.~Calabrese and J.~Cardy, \emph{Quantum quenches in extended systems},
  {\emph{Journal of Statistical Mechanics: Theory and Experiment} {\bfseries
  2007} (2007) P06008}.

\bibitem{Roberts:2014ifa}
D.~A. Roberts and D.~Stanford, \emph{{Two-dimensional conformal field theory
  and the butterfly effect}},
  \href{https://doi.org/10.1103/PhysRevLett.115.131603}{\emph{Phys. Rev. Lett.}
  {\bfseries 115} (2015) 131603}
  [\href{https://arxiv.org/abs/1412.5123}{{\ttfamily 1412.5123}}].

\bibitem{bazhanov1996integrable}
V.~V. Bazhanov, S.~L. Lukyanov and A.~B. Zamolodchikov, \emph{Integrable
  structure of conformal field theory, quantum kdv theory and thermodynamic
  bethe ansatz}, {\emph{Communications in Mathematical Physics} {\bfseries 177}
  (1996) 381}.

\bibitem{bazhanov1997integrable}
V.~V. Bazhanov, S.~L. Lukyanov and A.~B. Zamolodchikov, \emph{Integrable
  structure of conformal field theory ii. q-operator and ddv equation},
  {\emph{Communications in Mathematical Physics} {\bfseries 190} (1997) 247}.

\bibitem{bazhanov1999integrable}
V.~V. Bazhanov, S.~L. Lukyanov and A.~B. Zamolodchikov, \emph{Integrable
  structure of conformal field theory iii. the yang--baxter relation},
  {\emph{Communications in mathematical physics} {\bfseries 200} (1999) 297}.

\bibitem{bazhanov97zero}
V.~V. Bazhanov, S.~L. Lukyanov and A.~B. Zamolodchikov, \emph{Quantum field
  theories in finite volume: Excited state energies}, {\emph{Nuclear Physics B}
  {\bfseries 489} (1997) }.

\bibitem{maloney2018thermal}
A.~Maloney, G.~S. Ng, S.~F. Ross and I.~Tsiares, \emph{Thermal correlation
  functions of kdv charges in 2d cft}, {\emph{Journal of High Energy Physics}
  {\bfseries 2019} (2019) }.

\bibitem{Kotousov:2019nvt}
G.~A. Kotousov and S.~L. Lukyanov, \emph{{Spectrum of the reflection operators
  in different integrable structures}},
  \href{https://arxiv.org/abs/1910.05947}{{\ttfamily 1910.05947}}.

\bibitem{LeFloch:2019wlf}
B.~Le~Floch and M.~Mezei, \emph{{KdV charges in $T\bar{T}$ theories and new
  models with super-Hagedorn behavior}},
  \href{https://doi.org/10.21468/SciPostPhys.7.4.043}{\emph{SciPost Phys.}
  {\bfseries 7} (2019) 043} [\href{https://arxiv.org/abs/1907.02516}{{\ttfamily
  1907.02516}}].

\bibitem{vidmar2016generalized}
L.~Vidmar and M.~Rigol, \emph{Generalized gibbs ensemble in integrable lattice
  models}, {\emph{Journal of Statistical Mechanics: Theory and Experiment}
  {\bfseries 2016} (2016) 064007}.

\bibitem{cardy2016quantum}
J.~Cardy, \emph{Quantum quenches to a critical point in one dimension: some
  further results}, {\emph{Journal of Statistical Mechanics: Theory and
  Experiment} {\bfseries 2016} (2016) 023103}.

\bibitem{de2016remarks}
J.~de~Boer and D.~Engelhardt, \emph{Remarks on thermalization in 2d cft},
  {\emph{Physical Review D} {\bfseries 94} (2016) 126019}.

\bibitem{maloney2018generalized}
A.~Maloney, G.~S. Ng, S.~F. Ross and I.~Tsiares, \emph{Generalized gibbs
  ensemble and the statistics of kdv charges in 2d cft}, {\emph{Journal of High
  Energy Physics} {\bfseries 2019} (2019) }.

\bibitem{GGE}
A.~Dymarsky and K.~Pavlenko, \emph{Generalized gibbs ensemble of 2d cfts at
  large central charge in the thermodynamic limit}, {\emph{Journal of High
  Energy Physics} {\bfseries 2019} (2019) 98}.

\bibitem{GGE2}
A.~Dymarsky and K.~Pavlenko, \emph{Exact generalized partition function of 2d
  cfts at large central charge}, {\emph{Journal of High Energy Physics}
  {\bfseries 2019} (2019) 77}.

\bibitem{Witten:1987ty}
E.~Witten, \emph{{Coadjoint Orbits of the Virasoro Group}},
  \href{https://doi.org/10.1007/BF01218287}{\emph{Commun. Math. Phys.}
  {\bfseries 114} (1988) 1}.

\bibitem{novikov1974periodic}
S.~P. Novikov, \emph{The periodic problem for the korteweg--de vries equation},
  {\emph{Funktsional'nyi Analiz i ego Prilozheniya} {\bfseries 8} (1974) 54}.

\bibitem{lazutkin1975normal}
V.~F. Lazutkin and T.~Pankratova, \emph{Normal forms and versal deformations
  for hill's equation}, {\emph{Functional Analysis and its applications}
  {\bfseries 9} (1975) 306}.

\bibitem{dubrovin1974periodicity}
B.~A. Dubrovin and S.~P. Novikov, \emph{A periodicity problem for the
  korteweg--de vries and sturm--liouville equations. their connection with
  algebraic geometry},  in \emph{Dokl. Akad. Nauk SSSR}, vol.~219,
  pp.~531--534, 1974.

\bibitem{novikov1984theory}
S.~Novikov, S.~Manakov, L.~Pitaevskii and V.~E. Zakharov, \emph{Theory of
  solitons: the inverse scattering method}. Springer Science \& Business Media,
  1984.

\bibitem{magri1978simple}
F.~Magri, \emph{A simple model of the integrable hamiltonian equation},
  {\emph{Journal of Mathematical Physics} {\bfseries 19} (1978) 1156}.

\bibitem{gervais1982dual}
J.-L. Gervais and A.~Neveu, \emph{Dual string spectrum in polyakov's
  quantization (ii). mode separation}, {\emph{Nuclear Physics B} {\bfseries
  209} (1982) 125}.

\bibitem{gel1975asymptotic}
I.~M. Gel'fand and L.~A. Dikii, \emph{Asymptotic behaviour of the resolvent of
  sturm-liouville equations and the algebra of the korteweg-de vries
  equations}, {\emph{Russian Mathematical Surveys} {\bfseries 30} (1975) 77}.

\bibitem{gervais1985infinite}
J.-L. Gervais, \emph{Infinite family of polynomial functions of the virasoro
  generators with vanishing poisson brackets}, {\emph{Physics Letters B}
  {\bfseries 160} (1985) 277}.

\bibitem{gervais1985transport}
J.-L. Gervais, \emph{Transport matrices associated with the virasoro algebra},
  {\emph{Physics Letters B} {\bfseries 160} (1985) 279}.

\bibitem{dymarsky2019zero}
A.~Dymarsky, K.~Pavlenko and D.~Solovyev, \emph{Zero modes of local operators
  in 2d cft on a cylinder}, {\emph{arXiv preprint arXiv:1912.13444} (2019) }.

\bibitem{perez2016boundary}
A.~P{\'e}rez, D.~Tempo and R.~Troncoso, \emph{Boundary conditions for general
  relativity on ads3 and the kdv hierarchy}, {\emph{Journal of High Energy
  Physics} {\bfseries 2016} (2016) 103}.

\bibitem{Fuentealba:2017omf}
O.~Fuentealba, J.~Matulich, A.~Pérez, M.~Pino, P.~Rodríguez, D.~Tempo et~al.,
  \emph{Integrable systems with bms$\_{3}$ poisson structure and the dynamics
  of locally flat spacetimes},
  \href{https://doi.org/10.1007/JHEP01(2018)148}{\emph{JHEP} {\bfseries 01}
  (2018) 148} [\href{https://arxiv.org/abs/1711.02646}{{\ttfamily
  1711.02646}}].

\bibitem{Ojeda:2019xih}
E.~Ojeda and A.~Pérez, \emph{Boundary conditions for general relativity in
  three-dimensional spacetimes, integrable systems and the kdv/mkdv
  hierarchies}, \href{https://doi.org/10.1007/JHEP08(2019)079}{\emph{JHEP}
  {\bfseries 08} (2019) 079}
  [\href{https://arxiv.org/abs/1906.11226}{{\ttfamily 1906.11226}}].

\bibitem{Brown:1986nw}
J.~D. Brown and M.~Henneaux, \emph{{Central Charges in the Canonical
  Realization of Asymptotic Symmetries: An Example from Three-Dimensional
  Gravity}}, \href{https://doi.org/10.1007/BF01211590}{\emph{Commun. Math.
  Phys.} {\bfseries 104} (1986) 207}.

\bibitem{Bunster:2014mua}
C.~Bunster, M.~Henneaux, A.~Perez, D.~Tempo and R.~Troncoso, \emph{{Generalized
  Black Holes in Three-dimensional Spacetime}},
  \href{https://doi.org/10.1007/JHEP05(2014)031}{\emph{JHEP} {\bfseries 05}
  (2014) 031} [\href{https://arxiv.org/abs/1404.3305}{{\ttfamily 1404.3305}}].

\bibitem{banados1992black}
M.~Banados, C.~Teitelboim and J.~Zanelli, \emph{Black hole in three-dimensional
  spacetime}, {\emph{Physical Review Letters} {\bfseries 69} (1992) 1849}.

\bibitem{GETH}
A.~Dymarsky and K.~Pavlenko, \emph{Generalized eigenstate thermalization
  hypothesis in 2d conformal field theories}, {\emph{Physical review letters}
  {\bfseries 123} (2019) 111602}.

\bibitem{Banerjee:2018tut}
S.~Banerjee, J.-W. Brijan and G.~Vos, \emph{{On the universality of late-time
  correlators in semi-classical 2d CFTs}},
  \href{https://doi.org/10.1007/JHEP08(2018)047}{\emph{JHEP} {\bfseries 08}
  (2018) 047} [\href{https://arxiv.org/abs/1805.06464}{{\ttfamily
  1805.06464}}].

\bibitem{Vos:2018vwv}
G.~Vos, \emph{{Vacuum block thermalization in semi-classical 2d CFT}},
  \href{https://doi.org/10.1007/JHEP02(2019)022}{\emph{JHEP} {\bfseries 02}
  (2019) 022} [\href{https://arxiv.org/abs/1810.03630}{{\ttfamily
  1810.03630}}].

\bibitem{Erices:2019onl}
C.~Erices, M.~Riquelme and P.~Rodríguez, \emph{Btz black hole with
  korteweg–de vries-type boundary conditions: Thermodynamics revisited},
  \href{https://doi.org/10.1103/PhysRevD.100.126026}{\emph{Phys.Rev.D}
  {\bfseries 100} (2019) 126026}
  [\href{https://arxiv.org/abs/1907.13026}{{\ttfamily 1907.13026}}].

\bibitem{spohn2019generalized}
H.~Spohn, \emph{Generalized gibbs ensembles of the classical toda chain},
  {\emph{Journal of Statistical Physics} (2019) 1}.

\bibitem{bulchandani1905gge}
V.~Bulchandani, X.~Cao and H.~Spohn, \emph{The gge averaged currents of the
  classical toda chain}, {\emph{arXiv preprint arXiv:1905.04548} }.

\bibitem{castro2016emergent}
O.~A. Castro-Alvaredo, B.~Doyon and T.~Yoshimura, \emph{Emergent hydrodynamics
  in integrable quantum systems out of equilibrium}, {\emph{Physical Review X}
  {\bfseries 6} (2016) 041065}.

\bibitem{ilievski2017ballistic}
E.~Ilievski and J.~De~Nardis, \emph{Ballistic transport in the one-dimensional
  hubbard model: The hydrodynamic approach}, {\emph{Physical Review B}
  {\bfseries 96} (2017) 081118}.

\bibitem{piroli2017transport}
L.~Piroli, J.~De~Nardis, M.~Collura, B.~Bertini and M.~Fagotti, \emph{Transport
  in out-of-equilibrium xxz chains: Nonballistic behavior and correlation
  functions}, {\emph{Physical Review B} {\bfseries 96} (2017) 115124}.

\bibitem{bastianello2018generalized}
A.~Bastianello, B.~Doyon, G.~Watts and T.~Yoshimura, \emph{Generalized
  hydrodynamics of classical integrable field theory: the sinh-gordon model},
  {\emph{SciPost Phys} {\bfseries 4} (2018) 33}.

\bibitem{doyon2019generalized}
B.~Doyon, \emph{Generalized hydrodynamics of the classical toda system},
  {\emph{Journal of Mathematical Physics} {\bfseries 60} (2019) 073302}.

\bibitem{alekseev1989path}
A.~Alekseev and S.~Shatashvili, \emph{Path integral quantization of the
  coadjoint orbits of the virasoro group and 2-d gravity}, {\emph{Nuclear
  Physics B} {\bfseries 323} (1989) 719}.

\bibitem{maldacena2016remarks}
J.~Maldacena and D.~Stanford, \emph{Remarks on the sachdev-ye-kitaev model},
  {\emph{Physical Review D} {\bfseries 94} (2016) 106002}.

\bibitem{stanford2017fermionic}
D.~Stanford and E.~Witten, \emph{Fermionic localization of the schwarzian
  theory}, {\emph{Journal of High Energy Physics} {\bfseries 2017} (2017) 8}.

\bibitem{callebaut2019entanglement}
N.~Callebaut and H.~Verlinde, \emph{Entanglement dynamics in 2d cft with
  boundary: Entropic origin of jt gravity and schwarzian qm}, {\emph{Journal of
  High Energy Physics} {\bfseries 2019} (2019) 45}.

\end{thebibliography}\endgroup

\end{document}